\documentclass{iau} 
\usepackage{graphicx}

\title[IAUS 334.~~Rediscovering our Galaxy] 
{Impact of NLTE on determinations of atmospheric parameters and chemical abundances of very metal-poor stars}

\author[Lyudmila Mashonkina et al.]   
{Lyudmila Mashonkina, Tatyana Sitnova, Yuri Pakhomov \and \\ Tatyana Ryabchikova}

\affiliation{Institute of Astronomy, Russian Academy of Sciences \\ Pyatnitskaya st. 48, RU-119017 Moscow, Russia \\ email: {\tt lima@inasan.ru} \\
}

\pubyear{2017}
\volume{334}  
\jname{Rediscovering our Galaxy}
\editors{C. Chiappini, I. Minchev, E. Starkenburg \& M. Valentini, eds.}
\begin{document}

\maketitle

\begin{abstract}
Based on high-resolution spectral observations for a sample of very metal-poor stars, we investigate how well stellar chemical abundances can be derived with available theoretical methods and computational tools.

\end{abstract}


Chemical abundances of very metal-poor (VMP) stars provide important clues for learning the
initial mass function in the early stages of galaxy formation, stellar nucleosynthesis sites, mixing of the SNe ejecta in the interstellar medium. One aims therefore to push the accuracy of the abundance analysis to the point where the trends of the stellar abundance ratios with metallicity can be robustly discussed.

The first step to this goal is a determination of a homogeneous set of stellar atmosphere parameters: effective temperature, $T_{\rm eff}$, surface gravity, log~$g$, Fe abundance (metallicity), [Fe/H], and microturbulence velocity, $\xi_t$. Having worked on a sample of 36 VMP stars in the dwarf spheroidal galaxies (dSphs) and 23 Milky Way (MW) halo giants, \cite[Mashonkina \etal\ (2017a)]{Mashonkina2017p} recommend (i) to derive $T_{\rm eff}$ from photometric methods, (ii) to attain log~$g$ from the star distances, wherever available; if not, the Fe~I/Fe~II ionisation equilibrium based on the non-local thermodynamic equilibrium (NLTE) has proven to be a robust alternative at [Fe/H] $\succsim -3.7$, (iii) to calculate [Fe/H] from the Fe~II lines, (iv) to check $T_{\rm eff}$ and log~$g$ with theoretical evolutionary tracks.

\begin{figure}[b]
\begin{center}
 \includegraphics[width=2.6in]{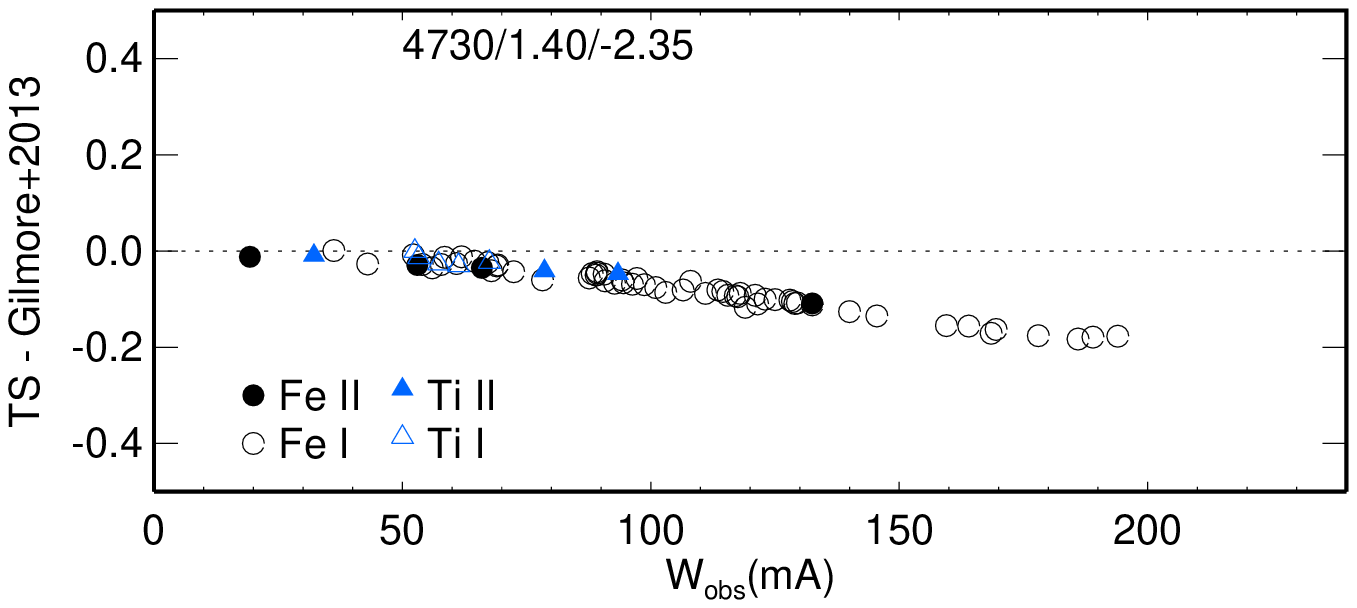}
 \includegraphics[width=2.6in]{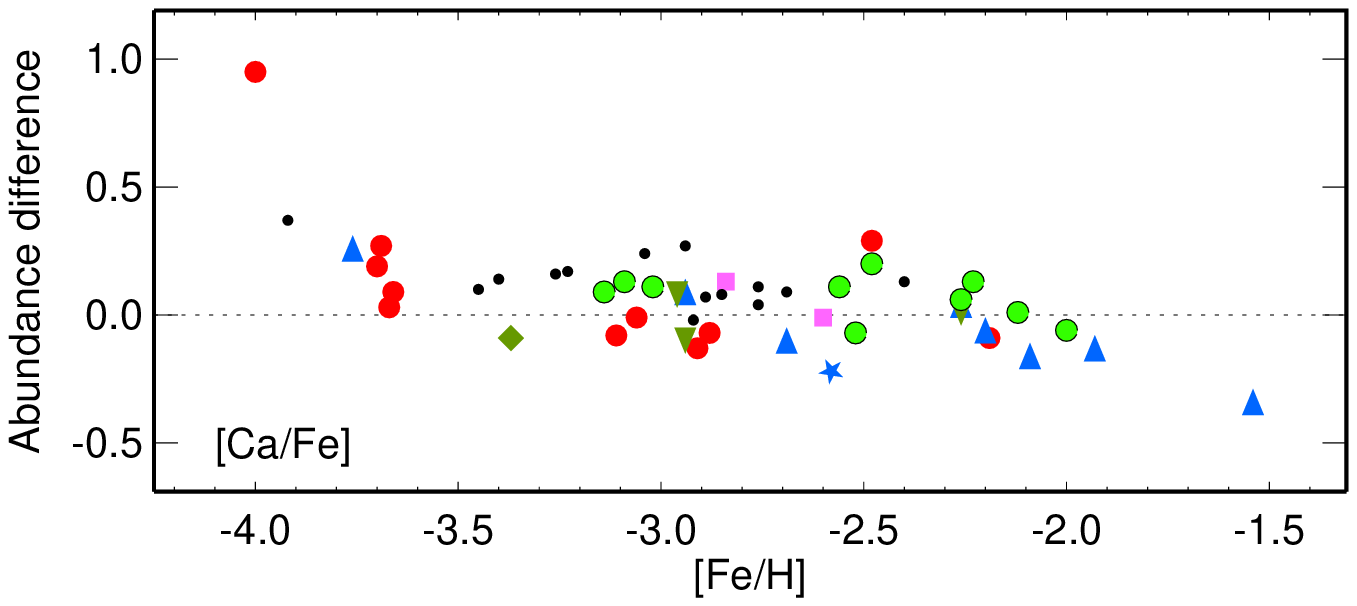}
 \caption{Left panel: differences in LTE abundances from lines of Fe~I, Fe~II, Ti~I, and Ti~II in Boo-33 between this study (TS) and GNM2013 (the NY analysis), when using common W$_{\rm obs}$s, $gf$-values, and model atmosphere parameters.
Right panel: differences in [Ca/Fe] between our NLTE and the other LTE studies for the stars in the dSphs Sculptor (circles, TJH2010, \cite[Jablonka \etal\ 2015]{Jablonka2015}, \cite[Simon \etal\ 2015]{Simon2015}, \cite[Kirby \& Cohen 2012: KC2012]{Kirby2012}), Ursa Minor (tilted triangles, \cite[Cohen \& Huang 2010]{Cohen2010}, \cite[Ural \etal\ 2015]{Ural2015}, KC2012), Sextans (squares, TJH2010), Fornax (rhombi, TJH2010), Bo\"otes~I (triangles, GNM2013, \cite[Norris \etal\ 2010]{Norris2010}, \cite[Frebel \etal\ 2016]{Frebel2016}), UMa~II (inverted triangles, \cite[Frebel \etal\ 2010]{Frebel2010}), and Leo~IV (5 pointed star, SFM10) and in the MW halo (small circles, CCT13).}
   \label{fig1}
\end{center}
\end{figure}

For that same stellar sample, we determined the LTE and NLTE abundances of up to 10 chemical elements \cite[(Mashonkina \etal, 2017b, Paper~I)]{Mashonkina2017a}, compared our results with the other studies, and explored the sources of abundance discrepancies. Figure~\ref{fig1} (left panel) displays the differences in the LTE abundances for the Ti and Fe lines in Boo-33, when using common observed equivalent widths (W$_{\rm obs}$), $gf$-values, $T_{\rm eff}$ = 4730~K, log~$g$ = 1.4, [Fe/H] = $-2.35$, and $\xi_t$ = 2.8~km\,s$^{-1}$ taken from \cite[Gilmore \etal\ (2013, GNM2013)]{Gilmore2013}. The abundance difference increases towards larger W$_{\rm obs}$ leading to suspect applying smaller van der Waals damping constants in GNM2013 compared with the most up-to-date $\Gamma_6$ values used in this study. Our suspicion was confirmed via a private communication with David Yong. As a result, we determine lower $\xi_t$ for Boo-33, by 0.5~km\,s$^{-1}$.

For [Ca/Fe], we inspect the combined effect of using different $T_{\rm eff}$, log~$g$, $\xi_t$, Fe abundance, and line-formation treatment in this NLTE and the other LTE studies (Fig.~\ref{fig1}, right panel). A clear outlier at [Fe/H] = $-4$ is Scl07-50, for which we used Ca~II 3933\,\AA, while  \cite[Tafelmeyer \etal\ (2010, TJH2010)]{Tafelmeyer2010} did Ca~I 4226\,\AA\ that leads to strongly underestimated element abundance, as discussed in Paper~I. When using lines of Ca~I and Fe~I, NLTE leaves [Ca/Fe] nearly unchanged compared with LTE because of similarly positive NLTE abundance corrections for lines of both chemical species. This explains small ($<$ 0.1~dex, in absolute value) differences in [Ca/Fe] for a part of our stellar sample, where corrections of the original atmospheric parameters were minor. \cite[Cohen \etal\ (2013, CCT13)]{Cohen2013} used lines of Ca~I and Fe~II under the LTE assumption. As a result, we obtain positive differences in [Ca/Fe] between this study and CCT13.
Negative differences in [Ca/Fe] for five [Fe/H] $> -2.7$ stars in Bo\"otes~I are caused by overestimating $\xi_t$ and underestimating the Fe abundance in GNM2013, as discussed above. For Leo IV-S1, our determination of the Fe abundance is 0.6~dex higher than in \cite[Simon \etal\ (2010, SFM10)]{Simon2010}, as a consequence of our higher $T_{\rm eff}$, by 200~K from the original estimate.
In the other cases, small negative differences in [Ca/Fe] are due to correction of the original atmospheric parameters. 

To summarize, a success of spectroscopic method in deriving the chemical abundances of stars is provided by not only advanced instrumentation and observations, but also careful determination of stellar atmosphere parameters, correct line-formation treatment, and using accurate atomic line data.

{\it Acknowledgements.}  L.M. and T.S. thank a support from the International Astronomical Union, the Russian Foundation for Basic Research (15-02-06046), and the grant on Leading Scientific
Schools 9951.2016.2 of the participation at IAUS334.


\smallskip
\begin{thebibliography}{}

\bibitem[Cohen \etal\ (2013)]{Cohen2013}
{Cohen, J.G., Christlieb, N., Thompson, I., \etal } 2013, \textit{ApJ}, 778, 56

\bibitem[Cohen \& Huang (2010)]{Cohen2010}
{Cohen, J.G. \& Huang, W.} 2010, \textit{ApJ}, 719, 931 

\bibitem[Frebel \etal\ (2010)]{Frebel2010}
{Frebel, A., Simon, J.D., Geha, M. \& Willman, B.} 2010, \textit{ApJ}, 708, 560             

\bibitem[Frebel \etal\ (2016)]{Frebel2016}
{Frebel, A., Norris, J.E., Gilmore, G. \& Wyse, R.F.G.} 2016, \textit{ApJ}, 826, 110

\bibitem[Gilmore \etal\ (2013)]{Gilmore2013}
{Gilmore, G., Norris, J.E., Monaco, L., \etal } 2013, \textit{ApJ}, 763, 61

\bibitem[Jablonka \etal\ (2015)]{Jablonka2015}
{Jablonka, P., North, P., Mashonkina, L., \etal } 2015, \textit{A\&A}, 583, A67

\bibitem[Kirby \& Cohen (2012)]{Kirby2012}
{Kirby, E. \& Cohen, J.G.} 2012, \textit{AJ}, 144, 108          

\bibitem[Mashonkina \etal\ (2017a)]{Mashonkina2017p}
{Mashonkina, L., Jablonka, P., Pakhomov, Y., \etal} 2017, \textit{AA}, in press

\bibitem[Mashonkina \etal\ (2017b)]{Mashonkina2017a}
{Mashonkina, L., Jablonka, P., Sitnova, T., \etal} 2017, \textit{AA}, submitted

\bibitem[Norris \etal\ (2010)]{Norris2010}
{Norris, J.E., Yong, D., Gilmore, G., \& Wyse, R.F.G.} 2010, \textit{ApJ}, 711, 350

\bibitem[Simon \etal\ (2010)]{Simon2010}
{Simon, J.D., Frebel, A., McWilliam, A., \etal} 2010, \textit{ApJ}, 716, 446             

\bibitem[Simon \etal\ (2015)]{Simon2015}
{Simon, J. D., Jacobson, H.R., Frebel, A., \etal} 2015, \textit{ApJ}, 802, 93

\bibitem[Tafelmeyer \etal\ (2010)]{Tafelmeyer2010}
{Tafelmeyer, M., Jablonka, P., Hill, V., \etal } 2010, \textit{A\&A}, 524, A58

\bibitem[Ural \etal\ 2015]{Ural2015}
{Ural, U., Cescutti, G., Koch, A., \etal} 2015, \textit{MNRAS}, 449, 761
\end{thebibliography}
\end{document}